\documentclass[intlimits,twoside,a4paper]{article}
\usepackage{amsmath,amssymb}
\usepackage{graphicx}
\usepackage[T2A]{fontenc}
\usepackage[cp1251]{inputenc}

\usepackage[eqsecnum]{cmpj}

\issue{2011}{14}{1}{13002}

\doinumber{10.5488/CMP.14.13002}

\title[Phase transitions of geometrically frustrated Ising-Heisenberg model]%
{Phase transitions of geometrically frustrated mixed
spin-1/2 and spin-1 Ising-Heisenberg model\\
 on diamond-like decorated planar lattices%
}
\author[L.~G\'alisov\'a, J.~Stre\v{c}ka]{Lucia G\'alisov\'a\refaddr{label1},
        Jozef Stre\v{c}ka\refaddr{label2}}
\addresses{
\addr{label1} Department of Applied Mathematics, Faculty of
Mechanical Engineering, Technical University, Letn\'a~9,
04200 Ko\v{s}ice, Slovak Republic
\addr{label2} Department of Theoretical Physics and Astrophysics,
Faculty of Science, P.~J.~\v{S}af\'arik University, Park
Angelinum~9, 04001 Ko\v{s}ice, Slovak Republic }

\date{Received July 26, 2010, in final form October 21, 2010}
\authorcopyright{L.~G\'alisov\'a, J.~Stre\v{c}ka}

\begin{document}

\maketitle

\begin{abstract}
Phase transitions of the mixed spin-1/2 and spin-1
Ising-Heisenberg model on several decorated planar lattices
consisting of interconnected diamonds are investigated within the
framework of the generalized de\-coration-iteration
transformation. The main attention is paid to the systematic study
of the finite-temperature phase diagrams in dependence on the
lattice topology. The critical behaviour of the hybrid
quantum-classical Ising-Heisenberg model is compared with the
relevant behaviour of its semi-classical Ising analogue. It is
shown that both models on diamond-like decorated planar lattices
exhibit a striking critical behaviour including reentrant phase
transitions. The higher the lattice coordination number is, the
more pronounced reentrance may be detected.
\keywords Ising-Heisenberg model, decoration-iteration
transformation, geometric frustration, reentrant phase transitions
\pacs 05.50.+q, 75.10.Hk, 75.10.Jm, 68.35.Rh
\end{abstract}

\section{Introduction}
\label{Intro}

Low-dimensional quantum spin models provide an excellent
playground for theoretical studies of cooperative and quantum
phenomena. It should be pointed out that quantum spin systems,
which are prone to a mutual interplay of geometric frustration,
thermal and quantum fluctuations, reveal the most remarkable
properties~\cite{Lhu02,Rich04,Mis04}. Under this specific
constraint, one often encounters a strikingly rich physics
manifested in diverse unusual ordered and disordered ground
states~\cite{Lhu02}, the order-from-disorder
effect~\cite{Lhu02,Rich04,Mis04}, the
chirality~\cite{Vil77,Vil80}, the enhanced magnetocaloric
effect~\cite{Zhi03,Hon05,Der06,Sch07} or the quantized
magnetization plateaus observable at low enough
temperatures~\cite{Rich04,Hon04}.

It is also well-known dictum that the theoretical investigation of
quantum spin models is frequently accompanied by rather
cumbersome and sophisticated mathematics, which precludes an exact
treatment of the most (even simple-minded) spin systems. Hence, it
follows that one usually has to rely on an application of some
approximative method(s) when treating the vast majority of
low-dimensional quantum spin models. However, the approximative
methods might have a profound deficiency in that they need not be
reliable enough in order to decide whether the observed (often
subtle) quantum phenomena are of a real physical significance or
they merely arose as an artefact of the applied approximation(s).
From this point of view, it is highly desirable to search for
artificial but exactly tractable quantum spin models, which might
display non-trivial quantum effects without a danger of
over-interpretation inherent in any approximation.

The hybrid {\it Ising-Heisenberg models on diamond-like decorated
lattices}~\cite{Str02,Can04,Str06,Can06,Can08,Jas08,Str09,Can09,Can10}
belong to the simplest exactly solved quantum spin models, which
were envisaged for describing lattice-statistical systems composed
of the semi-classical Ising and the quantum Heisenberg spins.
These simplified quantum-classical models can be rigorously
treated within the framework of generalized decoration-iteration
mapping transformation~\cite{Fis59,Syo72,Roj09}, because the nodal
Ising spins represent a barrier for quantum fluctuations that are
consequently restricted to elementary diamond-shaped units only.
It is also worth mentioning that the Ising-Heisenberg models on
the diamond-like decorated lattices have turned out to be a very
useful testing ground for elucidating several typical quantum
features.  Indeed, these interesting but still exactly tractable
spin systems may exhibit diverse quantum ordered and disordered
ground states~\cite{Str02,Can04,Str06,Can06,Can08,Can09,Can10},
the multi-step magnetization process with quantized intermediate
magnetization plateaus~\cite{Can06,Str09,Can09}, the enhanced
magnetocaloric effect~\cite{Can06,Can09}, as well as the
non-trivial criticality~\cite{Str06,Jas08,Can10}.

In our recent works~\cite{Str02,Jas08,Can10}, we have examined in
detail the ferromagnetic mixed spin-1/2 and spin-1
Ising-Heisenberg model on several diamond-like decorated planar
lattices. It has been demonstrated that this hybrid
classical-quantum model exhibits a rather rich ground-state and
finite-temperature phase diagrams on account of the competition
between the easy-axis Ising and the easy-plane Heisenberg
interaction. The main purpose of the present work is to
investigate critical properties of the antiferromagnetic mixed
spin-1/2 and spin-1 Ising-Heisenberg model on diamond-like
decorated planar lattices, which might even display more complex
and non-trivial criticality owing to an interplay of the geometric
frustration, thermal and quantum fluctuations as evidenced by our
preliminary ground-state analysis~\cite{Can08}.

The organization of this paper is as follows. In the next section,
we shall define the mixed spin-1/2 and spin-1 Ising-Heisenberg
model on diamond-like decorated lattices and we shall also briefly
review the most important steps of an exact calculation based on
the decoration-iteration procedure. The section~\ref{Sec3} deals
with a discussion of the most interesting numerical results for
finite-temperature phase diagrams and temperature variations of
the total magnetization both for the classical-quantum
Ising-Heisenberg model and its semi-classical Ising
analogue. In this section, our main attention will be focused on
a possibility of observing reentrant phenomenon.
Finally, some concluding remarks are drawn in section~\ref{Sec4}.

\section{Formulation}
\label{Sec2}

\begin{figure}[!hd]
 \vspace*{-3.25cm}
\begin{center}
\includegraphics[width= 0.8\textwidth]{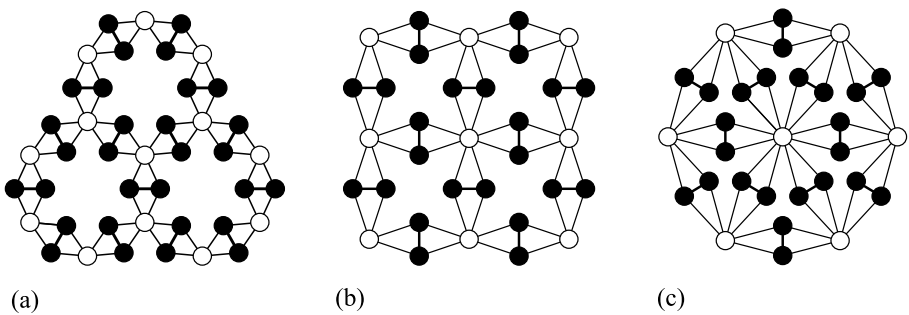}
  \vspace*{-2.5cm}
\caption{The mixed spin-$1/2$ and spin-$1$ Ising-Heisenberg model
on diamond-like decorated honeycomb (figure~\ref{fig1}a), square
(figure~\ref{fig1}b) and triangular (figure~\ref{fig1}c) lattices.
White circles represent the nodal Ising spins $\sigma = 1/2$ and
black ones denote the decorating Heisenberg spins $S = 1$.}
\label{fig1}
\end{center}
\vspace*{-0.5cm}
\end{figure}

Let us consider two-dimensional lattices composed of
inter-connected diamonds as is illustrated in figure~\ref{fig1}
for honeycomb, square and triangular lattices. In this figure, the
empty circles denote lattice positions of the Ising spins
$\sigma=1/2$ that interact with other spins through the
interaction $J_{\rm I}$ and the full ones represent lattice
positions of the decorating Heisenberg spins $S=1$ that interact
among themselves via the anisotropic XXZ coupling $J_{\rm
H}(\Delta)$. Note that the parameter $\Delta$ allows to control
the interaction $J_{\rm H}$ between the easy-axis ($\Delta<1$) and
easy-plane ($\Delta>1$) type, as well as, to obtain the Ising
model as a special limiting case when assuming $\Delta=0$.
Providing that the parameter $D$ stands for the axial zero-field
splitting (AZFS) parameter~\cite{Rud08a,Rud08b}, which acts on the
decorating Heisenberg spins $S=1$ only,
the total Hamiltonian of the investigated model can be written as
\begin{eqnarray}
\hat{{\cal H}} = - J_{\rm H}\sum_{(i,j)}^{Nq/2}\,
[\Delta(\hat{S}_{i}^{x}\hat{S}_{j}^{x}+\hat{S}_{i}^{y}\hat{S}_{j}^{y})
+\hat{S}_{i}^{z}\hat{S}_{j}^{z}] - J_{\rm I}\sum_{(i,n)}^{2Nq}
\hat{S}_{i}^{z}\hat{\sigma}_{n}^{z} - D\sum_{i = 1}^{Nq}\,
(\hat{S}_{i}^{z})^{2}. \label{eq:Hih}
\end{eqnarray}
Here, the first summation is carried out over all interactions
between pairs of the nearest-neigh\-bouring Heisenberg spins, the
second summation takes into account the interaction between the
nearest-neighbouring Ising and Heisenberg spins and the last summation
runs over all lattice sites occupied by the decorating Heisenberg
spins. The spin variables $\hat{\sigma}_{n}^{z}$ and
$\hat{S}_{i}^{\gamma}$ ($\gamma = x, y, z$) denote spatial
components of the spin-$1/2$ and spin-$1$ operators located at
$n$th and $i$th lattice position, respectively. Finally, $N$
represents the total number of the Ising spins, and $q$ is the
coordination number of the lattice, which implies that the total
number of all spins (lattice sites) is $N_{\mathrm{tot}} =
N(q+1)$. It is worth mentioning that the quantum-classical model defined in this way is exactly solvable within the framework of
a generalized decoration-iteration mapping
transformation~\cite{Fis59,Syo72,Roj09} (for more computational
details see our recent work~\cite{Can10}, where we presented a
detailed procedure for an exact solution of this model). As a
result of this mapping, one obtains a simple relation between the
partition function ${\cal Z}$ of the investigated mixed spin-$1/2$
and spin-$1$ Ising-Heisenberg model on diamond-like decorated
lattices and the partition function ${\cal Z}_{\rm Ising}$ of the
simple spin-$1/2$ Ising model on corresponding undecorated
lattices with the nearest-neighbour coupling $R$:
\begin{equation}
{\cal Z} (T, J_{\rm I} , J_{\rm H}, \Delta, D) = A^{Nq/2} {\cal
Z}_{\rm Ising}(T, R). \label{eq:Zih=Zi}
\end{equation}
Note that both mapping parameters $A$ and $R$ are unambiguously
determined by a self-consistency condition of the applied
decoration-iteration mapping transformation and their explicit
forms are given by equations (5)--(7) of reference~\cite{Can10}.

At this stage, it is worthwhile to remark that the mapping
relation~(\ref{eq:Zih=Zi}) is universal and valid regardless of
the lattice topology and spatial dimensionality of the model
system. In addition, it also permits a rather comprehensive
analysis of its critical behaviour as well as some basic
thermodynamic quantities. Indeed, it directly follows from
equation~(\ref{eq:Zih=Zi}) that the investigated mixed-spin
Ising-Heisenberg model becomes critical if and only if the
spin-$1/2$ Ising model with the effective coupling $R$ on the
corresponding undecorated lattice becomes critical as well.
Bearing this in mind, the exact critical temperature of the
spin-1/2 and spin-1 Ising-Heisenberg model on diamond-like
decorated honeycomb, square and triangular lattices can be
straightforwardly obtained from the conditions
\begin{eqnarray}
\label{eq:Tc} \beta_{\mathrm{c}}R = 2\ln(2+\sqrt{3})\,, \qquad
\beta_\mathrm{c}R = 2\ln(1+\sqrt{2})\,, \qquad \beta_\mathrm{c}R =
\ln 3,
\end{eqnarray}
that represent exact relationships between the effective
temperature-dependent coupling $\beta R$ and relevant critical
temperatures of the simple spin-$1/2$ Ising model on a corresponding
undecorated honeycomb~\cite{Hou50}, square~\cite{Ons44} and
triangular~\cite{Bax89} lattices, respectively. In the above,
$\beta_\mathrm{c}= 1/(k_\mathrm{B}T_\mathrm{c})$ and
$T_\mathrm{c}$ is the critical temperature of the studied quantum
mixed-spin Ising-Heisenberg model. Similarly, the sub-lattice and
total magnetization can also be derived from the exact mapping
equivalence (\ref{eq:Zih=Zi}) between the partition functions
${\cal Z}$ and ${\cal Z}_{\rm Ising}$. More specifically, by
combining equation~(\ref{eq:Zih=Zi}) with the exact mapping
theorems developed by Barry et al.~\cite{Bar88,Kha90,Bar91} and
the generalized Callen-Suzuki spin
identity~\cite{Cal63,Suz65,Bal02}, the sub-lattice
magnetization~$m_{\rm i}^{z}$,~$m_{\rm h}^z$ reduced per one Ising
and Heisenberg spin, respectively, can be directly computed from
the precise relations:
\begin{eqnarray}
\label{eq:miz}
m_{\rm i}^z \!\!\!\! &\equiv&\!\!\!\!
\langle \hat{\sigma}_{k1}^{z}
\rangle = \langle \hat{\sigma}_{k1}^{z} \rangle_{0} \equiv m_0,
\\
\label{eq:mhz}
m_{\rm h}^z\!\!\!\! &\equiv&\!\!\!\! \langle \hat{S}_{k1}^{z}\rangle =
4m_0\,F(\beta J_{\rm I}),
\end{eqnarray}
where $ F(x) = \left[\exp(2\beta D +\beta J_{\rm H})\sinh(2x) + \exp(\beta D)\sinh(x)\cosh(\beta J_{\rm H}\Delta)\right]/W_1\,$,
the parameter $W_1$ is defined by equation (6) of reference~\cite{Can10} and the symbols $\langle\ldots\rangle$ and $\langle\ldots\rangle_0$ represent standard canonical averages performed over the ensemble defined by the mixed-spin Ising-Heisenberg model on the diamond-like decorated lattice and the spin-$1/2$ Ising model on the corresponding lattice, respectively, and $m_0$ labels the single-site magnetization of the corresponding Ising model. In view of this notation, the total magnetization reduced per spin of the investigated lattice, which represents the order parameter of the mixed-spin Ising-Heisenberg model on diamond-like decorated lattices, can be expressed as follows: $m = (m_{\rm i}^z + qm_{\rm h}^z)/(q+1)$.

\section{Results and discussion}
\label{Sec3}

In this part, let us proceed to a discussion of the most
interesting results obtained for the mixed spin-$1/2$ and spin-$1$
Ising-Heisenberg model on diamond-like decorated honeycomb, square
and triangular lattices. Before doing this, however, it is worth
mentioning that all the results obtained in section~\ref{Sec2} are
universal since they hold true regardless of whether ferromagnetic or
antiferromagnetic interactions $J_{\rm I}$ and $J_{\rm H}$ are
assumed, and are independent of the lattice topology or
spatial dimensionality of the investigated system.

Moreover, the effective nearest-neighbour interaction $R$ of the
spin-1/2 Ising model on the corresponding undecorated lattice is
invariant under the transformation $J_{\rm I} \rightarrow - J_{\rm
I}$ [see equations~(5)--(7) in reference~\cite{Can10}]. This
observation leads to the conclusion that the critical temperature
as well as other thermodynamic quantities of the model under
investigation remain unchanged under the change of the nature of
the Ising interaction $J_{\rm I}$. Indeed, a change of the
ferromagnetic interaction $J_{\rm I}>0$ to the antiferromagnetic
one $J_{\rm I}<0$ causes just a trivial change in the local
alignment of the nodal Ising spins with respect to their nearest
Heisenberg neighbours. By contrast, there are some fundamental
differences between magnetic behaviour of models with distinct
nature of the Heisenberg interaction $J_{\rm H}(\Delta)$ (see our
preliminary reports~\cite{Str02,Can08,Jas08}). Taking
 this fact into account, we have restricted our recent work~\cite{Can10} to the
particular case with the ferromagnetic Heisenberg and Ising
interactions ($J_{\rm H}>0$, $J_{\rm I}>0$) only. Besides a rather
complex ground state composed of two unusual quantum phases, a
striking critical behaviour including reentrant phase transitions
with two or three consecutive critical points has been discussed
in this work. With this background, our attention
in this paper will be focused on the analysis of finite-temperature behaviour of
the mixed spin-$1/2$ and spin-$1$ Ising-Heisenberg model on
diamond-like decorated planar lattices, in which both the
interaction constants $J_{\rm I}$ and $J_{\rm H}$ are supposed to
be antiferromagnetic ($J_{\rm I}<0$, $J_{\rm H}<0$).

\subsection{Summary of preliminary results}

Before proceeding further, let us briefly summarize our previous
numerical results published in reference~\cite{Can08}. This work
was devoted to the comparison of the ground-state properties of
the quantum antiferromagnetic spin-$1/2$ and spin-$1$
Ising-Heisenberg model on diamond-like decorated planar lattices
and its semi-classical Ising analogue. The obtained results
revealed that the investigated model system has a rich ground-state
phase diagram, which consists of a semi-classically ordered
ferrimagnetic phase FRI$_{1}$ with the perfect antiparallel
alignment between the nearest-neighbouring Ising and Heisenberg
spins and two ferrimagnetic phases FRI$_{2}$ and FRI$_{3}$, which
differ one from another just in a quantum entanglement of the
Heisenberg spin pairs described by the antisymmetric wave function
$(|1,0\rangle - |0,1\rangle)/\sqrt{2}$ that emerges just in
FRI$_{2}$. The latter ferrimagnetic phase FRI$_{3}$ represents the
classical ferrimagnetic phase, where the decorating spin pairs may
reside in one of the two possible spin states: either $|1,0\rangle$ or
$|0,1\rangle$. According to this ambiguous order of the decorating
spins, one may regard the phase FRI$_{3}$ as a partly
degenerate state with the non-zero residual entropy $S_0/N_{\rm
tot}k_{\rm B} = [{q}/{(2q+2)}]\ln2$, which is proportional to the
coordination number $q$ of the lattice (total number of decorating
spin pairs). Finally, several geometrically frustrated phases can
also be found in the ground state depending on whether the hybrid
quantum-classical Ising-Heisenberg model or its semi-classical
Ising version is considered. In the former case, the ground state
is formed by a frustrated phase FRU, where all nodal Ising spins
are frustrated due to a quantum superposition of three spin states
$|0,0\rangle$, $|1,-1\rangle$ and $|-1,1\rangle$ of the decorating
Heisenberg spins, whose relative probabilities depend on a mutual
ratio between interaction parameters. On the other hand, two
different frustrated phases FRU$_{1}$ and FRU$_{2}$ can be
detected for the semi-classical Ising analogue of the model. In
the former phase FRU$_{1}$\,, the geometric frustration of the
nodal Ising spins is caused by ``non-magnetic'' nature of the
Heisenberg spin dimers $|0,0\rangle$, while the geometric
frustration in the latter phase FRU$_{2}$\, comes from
antiferromagnetic spin states (either $|1,-1\rangle$ or
$|-1,1\rangle$) of the Heisenberg spin pairs. As it has been shown
in reference~\cite{Can08}, both these phases can be regarded as
special limiting cases of the unique frustrated phase FRU.

\subsection{Finite-temperature behaviour of the semi-classical Ising model}

\begin{figure}
\begin{center}
\includegraphics[width= 0.8\textwidth]{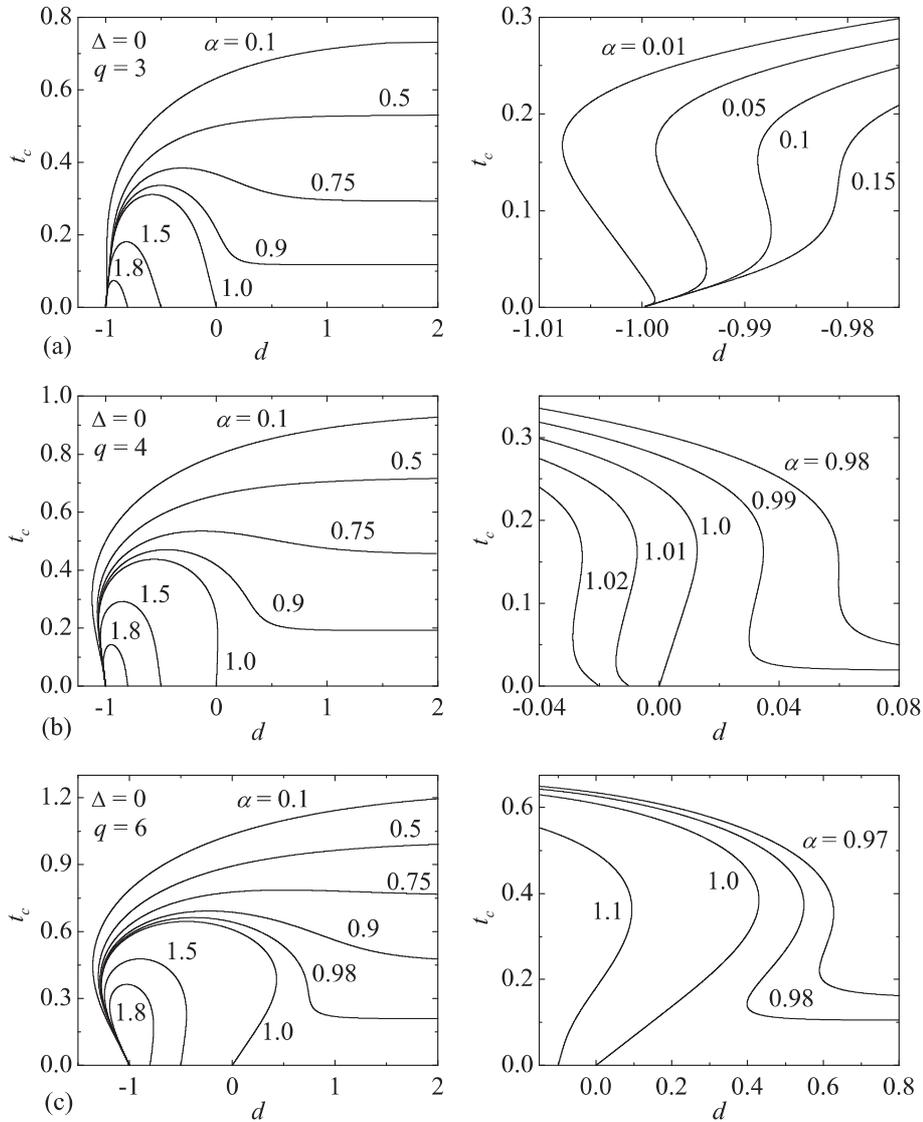}
\caption{The critical temperature of the
semi-classical Ising limit of the mixed spin-$1/2$ and spin-$1$
Ising-Heisenberg model on diamond-like decorated honeycomb
(figure~\ref{fig2}a), square (figure~\ref{fig2}b) and triangular
(figure~\ref{fig2}c) lattices as a function of the AZFS parameter
$d$ for several values of the interaction ratio~$\alpha$. The
right column illustrates in detail the regions, where the system
exhibits reentrant phase transitions with two or three consecutive
critical points.} \label{fig2}
\end{center}
\end{figure}

Now, let us proceed to the discussion of the finite-temperature
behaviour of the antiferromagnetic spin-$1/2$ and spin-$1$
Ising-Heisenberg model on diamond-like decorated planar lattices.
To enable a direct comparison with the ground-state analysis
published in reference~\cite{Can08}, we start first with the
discussion of finite-temperature phase diagrams, which are
displayed in figure~\ref{fig2} in the reduced units $t = k_{\rm
B}T/|J_{\rm I}|$, $d = D/|J_{\rm I}|$ and $\alpha = |J_{\rm
H}|/|J_{\rm I}|$, describing the dimensionless temperature, the
relative strength of the AZFS parameter and the strength of the
Heisenberg interaction normalized with respect to the Ising
interaction, respectively. This figure shows the critical
temperature of the semi-classical Ising version of the
investigated mixed-spin model on diamond-like decorated honeycomb
(figure~\ref{fig2}a), square (figure~\ref{fig2}b) and triangular
(figure~\ref{fig2}c) lattices as a function of the AZFS parameter
$d$ for several values of the interaction ratio~$\alpha$. Note
that solid lines depicted in this figure are the unique solutions
of the critical conditions~(\ref{eq:Tc}) and, as a consequence,
 they represent the lines of the second-order phase
transitions separating the spontaneously ordered phases (${\rm
FRI}_1$ or~${\rm FRI}_3$) from the disordered paramagnetic one. As
one can clearly see from figure~\ref{fig2}, the overall critical
behaviour of the system very sensitively depends on the strength of
the interaction parameters $\alpha$ and $d$, as well as, the
topology (coordination number) of the lattice; for $\alpha<1$, the
critical temperature $t_{\rm c}$ either monotonously decreases
upon decrease of the AZFS parameter until it tends towards zero
temperature at the boundary value $d=-1$ (see the curves labeled
as $\alpha=0.1$ and~$0.5$), or it exhibits an interesting
non-monotonous dependence to be closely related to the ${\rm
FRI}_1\rightarrow{\rm FRI}_3$ phase transition when the
interaction ratio $\alpha$ is sufficiently close to the value
$\alpha=1$ (see e.g. the case $J_{\rm H}/J_{\rm I} = 0.9$ and
figure~2(a) in reference~\cite{Can08} for clarity).
On the other hand, if one considers~$1\leqslant\alpha\leqslant2$,
the non-zero critical temperature may be observed just around the
interval of the AZFS parameters $d \in \left(-1, 1-\alpha\right)$,
where the partly degenerate ferrimagnetic phase~${\rm FRI}_3$
constitutes the ground state. Furthermore, several interesting
regions with reentrant phase transitions can also be observed in
finite-temperature phase diagrams shown in figure~\ref{fig2}. More
specifically, the diamond-like decorated honeycomb lattice
exhibits the reentrant behaviour with either two or three
consecutive critical points before the critical line tends to zero
temperature when the interaction ratio $\alpha$ acquires very
small values, i.e., when the antiferromagnetic interaction $J_{\rm
H}$ between decorating spins is much less than the competitive
antiferromagnetic interaction $J_{\rm I}$ between the nodal and
decorating spins (see the detail of figure~\ref{fig2}a). This
non-trivial behaviour quickly vanishes as the competitive
interaction $J_{\rm H}$ strengthens (see the cases labeled as
$\alpha = 0.01$, $0.05$ and~$0.1$), which suggests that it appears
just as a result of a mutual competition between the interaction
$J_{\rm I}$ favouring the antiferromagnetic arrangement of nodal
Ising spins with respect to their nearest spin neighbours and the
AZFS parameter $D<0$ that tends to lower the spin states
of decorating spins. By contrast, other two diamond-like decorated
square and triangular lattices exhibit the reentrant behaviour
in the left neighbourhood of the boundary value $d = -1$,
where the frustrated phase ${\rm FRU}_1$ constitutes the ground
state, but for any $\alpha \in \left(0, 2\right)$ (see
figures~\ref{fig2}b and c). In these parts of finite-temperature
phase diagrams, both the lattices start from the disordered ground
state ${\rm FRU}_1$ before entering the partly ordered phase ${\rm
FRI}_3$ at lower critical temperature $t_{{\rm c}1}$, whose
spontaneous order disappears due to strong thermal fluctuations at the
upper critical temperature $t_{{\rm c}2}$. Note that the observed
non-trivial behaviour fully corresponds to the condition that the
coexistence of the spin order and disorder in the ground state of
frustrated Ising systems leads to the occurrence of the reentrant
phenomenon at non-zero temperatures~\cite{Diep04}. However, this condition
is  only necessary but not sufficient for the occurrence
of the reentrant phenomenon in Ising systems, which can be
directly seen from the regions around the boundary points $d = 1 -
\alpha$ ($\alpha \geqslant 1$) and~$d = \alpha - 1$ ($\alpha < 1$)
in figures~\ref{fig2}b, c. Obviously, just the square and
triangular lattices show the reentrance in these parts of phase
diagrams (see details of figures~\ref{fig2}b and c). As the
detailed ground-state analysis has revealed, the ground-state
degeneracy of the ferrimagnetic phase ${\rm FRI}_3$ is not
high enough in order to cause the reentrant phenomenon in
the case of honeycomb lattice (remember that the residual entropy
of the honeycomb lattice is $S_0/N_{\rm tot}k_{\rm B}=0.2599$,
while the residual entropy of the square and triangular lattices
is $S_0/N_{\rm tot}k_{\rm B}=0.2773$ and $0.2971$, respectively).
\begin{figure}[htb]%
\vspace*{-0.8cm}
\begin{center}
\includegraphics[width= 0.85\textwidth]{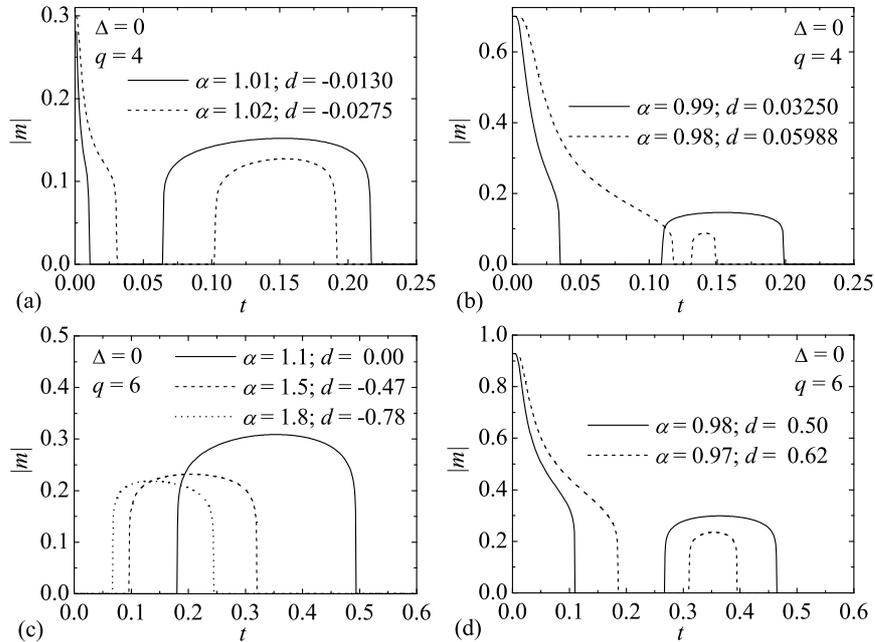}
\vspace*{-0.55cm}
\caption{Temperature dependencies of the total
magnetization of the decorated square (figures~\ref{fig3}a, b) and
triangular (figures~\ref{fig3}c, d) lattices for such combination
of the interaction parameters $\alpha$ and $d$, where the system
exhibits the reentrant behaviour.} \label{fig3}
\end{center}
\end{figure}
Actually, the square lattice exhibits the reentrant behaviour only
if competitive exchange interactions between the
nearest-neighbouring spins are equal or approximately equal to
each other (i.e.~if $\alpha \simeq 1$). On the other hand, the
triangular lattice shows the reentrance in this region for the
interaction ratio $\alpha \lesssim 1$ as well as for all $\alpha
\in \langle 1, 2)$ (see figure~\ref{fig2}c and its detail). It is
noteworthy that the region with reentrant phase transitions gradually
vanishes as the difference between the considered ratio $\alpha$
and the value $1$ increases. These results can also be convincingly
evidenced by thermal dependencies of the absolute value of the
total magnetization $|m|$ displayed in figure~\ref{fig3} for
diamond-like decorated square (figure~\ref{fig3}a, b) and
triangular (figure~\ref{fig3}c, d) lattices.

\subsection{Finite-temperature behaviour of the isotropic quantum version of the model}

\begin{figure}
\begin{center}
\includegraphics[width= 0.8\textwidth]{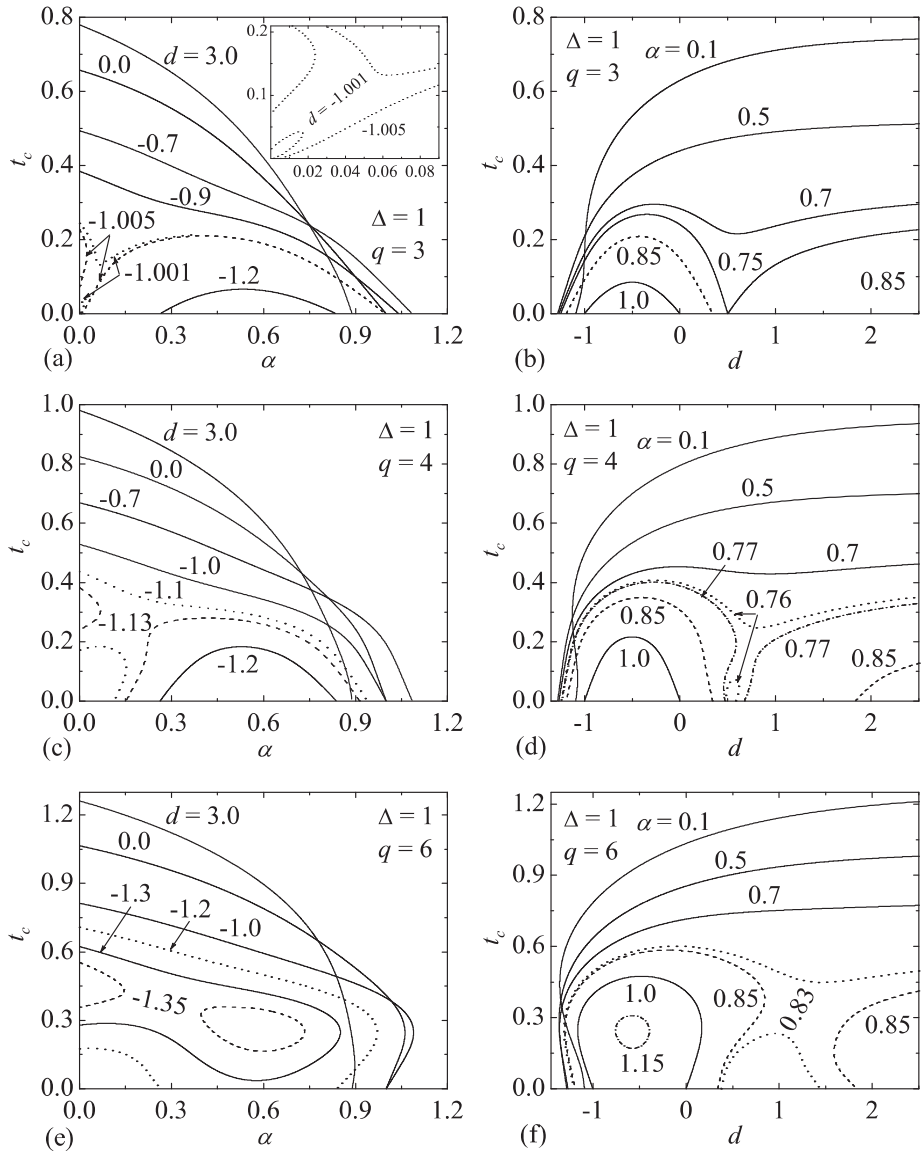}
\caption{The critical temperature of the mixed
spin-$1/2$ and spin-$1$ Ising-Heisenberg model on diamond-like
decorated honeycomb (figures~\ref{fig4}a, b), square
(figures~\ref{fig4}c, d) and triangular (figures~\ref{fig4}e, f)
lattices the fixed the exchange anisotropy $\Delta = 1$ as a
function of the interaction ratio $\alpha$ for several values of
the AZFS parameter $d$ and as a function of the AZFS parameter $d$
for several values of $\alpha$.} \label{fig4}
\end{center}
\end{figure}

To provide a deeper insight into the finite-temperature behaviour
of the quantum version of the investigated model system, let us
turn our attention to the phase diagrams displayed in
figure~\ref{fig4}. This figure shows the critical temperature of
the mixed spin-1/2 and spin-1 Ising-Heisenberg model on
diamond-like decorated honeycomb, square and triangular lattices
as a function of the interaction ratio $\alpha$ and the AZFS
parameter $d$ for the fixed the exchange anisotropy $\Delta = 1$.
Apart from the expected monotonous decrease of $t_{\rm c}$ with the
increasing ratio $\alpha$ and/or decreasing $d$, one may also find
here more interesting non-monotonic dependencies of $t_{\rm c}$ as
well as several striking regions with reentrant phase transitions.
More specifically, the honeycomb lattice shows the reentrant
behaviour with either two or three consecutive critical points
just for very small values of the interaction ratio $\alpha$ and
for the particular values of the AZFS paremeter $d\gtrsim-1$. It
is quite obvious that the origin of this non-trivial phenomenon
lies in a mutual competition between the antiferromagnetic Ising
interaction $J_{\rm I}$ and the easy-plane AZFS parameter $D<0$.
As could be expected, the parameter region with reentrant phase
transitions enlarges with the increasing coordination number of
the lattice. Indeed, this becomes quite clear from
figures~\ref{fig4}c and e, the spin-1/2 and spin-1
Ising-Heisenberg model on the decorated square and triangular
lattices exhibits two or three critical points for $D>0$ and
$J_{\rm H}(\Delta)$, whereas the effect of the former
interaction parameter is supported by the Ising interaction
$J_{\rm I}$. Finally, the remarkable parts of finite-temperature
phase diagrams represent particular dependencies of $t_{\rm c}$
forming closed loops (see figure~2 in reference~\cite{Hon04b}). In
these regions of phase diagrams, the system starts with the
disordered ground state before entering the spontaneously ordered
ferrimagnetic phase FRI$_2$ at lower critical temperature $t_{{\rm
c}1}$, which subsequently disappears due to strong thermal
fluctuations at upper critical temperature $t_{{\rm c}2}$ (see the
curves $d=-1.35$ in figure~\ref{fig4}e and
$\alpha=1.15$ in figure~\ref{fig4}f).
\begin{figure}[htb]%
\begin{center}
\includegraphics[width= 0.85\textwidth]{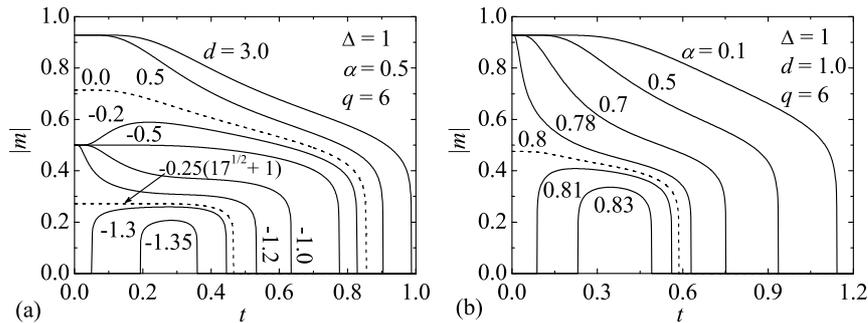}
\caption{Temperature dependencies of the total
magnetization of the mixed-spin Ising-Heisenberg model on
diamond-like decorated triangular lattice with the fixed the
exchange anisotropy $\Delta = 1$ for the interaction ratio
$\alpha=0.5$ and several values of the AZFS parameter $d$
(figure~\ref{fig5}a) and for the AZFS parameter $d=1$ and several
values of the ratio $\alpha$. The dependencies depicted as broken
curves correspond to the two-phase coexistence at $t=0$ (see the
discussion).} \label{fig5}
\end{center}
\end{figure}
The results displayed in figure~\ref{fig4} can be also convincingly
evidenced by thermal dependencies of the total magnetization.
Since the most diverse finite-temperature behaviour has been
observed for the diamond-like decorated triangular lattice, in figure~\ref{fig5} we
have depicted some typical temperature
variations of the absolute value of total magnetization $|m|$ for
this particular model system. As one can clearly see, the
magnetization exhibits reentrant transitions with two consecutive
critical points for interaction ratio $\alpha=0.5$ and the AZFS
parameters $d\lesssim-0.25(\sqrt{17}+1)$ as depicted in
figure~\ref{fig5}a or for the particular case of the AZFS
parameter $d=1$ and the interaction ratio $\alpha\gtrsim0.8$ as
depicted in figure~\ref{fig5}b. Note that these values of the
interaction parameters $\alpha$ and $d$ are from a close vicinity
of the ground-state phase transition between the ferrimagnetically
ordered phase (FRI$_1$ or FRI$_2$) and the disordered FRU phase.
On the other hand, if one considers the parameter space
$d>-0.25(\sqrt{17}+1)$ for the interaction ratio $\alpha=0.5$ or
$\alpha<0.8$ for the AZFS parameter $d=1$, then the displayed
magnetization curves $|m(t)|$ exhibit a single critical point only
(see, e.g., the curves $d=-0.5,\,0.5$ in figure~\ref{fig5}a and
$\alpha=0.1,\,0.5,\,0.7$ in figure~\ref{fig5}b).

\section{Concluding remarks}
\label{Sec4} In the present paper, the critical behaviour of the
mixed spin-$1/2$ and spin-$1$ Ising-Heisenberg model on
diamond-like decorated honeycomb, square and triangular lattices
has been investigated within the framework of the generalized
decoration-iteration mapping transformation. Using this rigorous
procedure, the exact solution for the investigated mixed-spin
model has been obtained by establishing a precise mapping
equivalence with the spin-$1/2$ Ising model on the corresponding
undecorated planar lattice with the known exact solution.

The main emphasis of this work was laid upon the systematic study
of the finite-temperature phase diagrams that basically depend on
the lattice topology and upon the comparison of the critical
behaviour of the hybrid quantum-classical Ising-Heisenberg model
and its semi-classical Ising variant. The most interesting result
consists in the exact evidence of the existence of reentrant phase
transitions with either two or three consecutive critical points.
The presence of this non-trivial phenomenon has also been
evidenced by temperature dependencies of the total magnetization.
It has been proved that the higher the coordination number of the
lattice is, the more pronounced reentrance is and the more diverse
critical behaviour may be found.

Finally, it is worthwhile to remark that even though our
theoretical investigation of the mixed spin-$1/2$ and spin-$1$
Ising-Heisenberg model on diamond-like decorated planar lattices
has been mainly aimed at providing a deeper insight into
cooperative and quantum features of this exactly solvable model,
we hope that our results might stimulate research on possible
experimental realizations of this interesting quantum spin model
and confirm our theoretical predictions. From this perspective,
the most promising approach in experimental realizations of our
model system might represent a targeted design of cyano-based
polymeric coordination compounds or their isostructural analogues.
For instance, the series of bimetallic polymeric coordination
compounds $\rm \{Cu(L)\}_3[Fe(CN)_6](ClO_4)_2\cdot nH_2O$, L =
N-(3\,-aminopropyl)-1,\,3\,-propanediamine~\cite{Zha00,Hon04b} or
N-(2\,-aminoethyl)-1,\,3\,-propanediamine~\cite{Tra01}, which
possess the diamond-like decorated honeycomb network structure,
may represent a useful starting point for this rational
synthesis. In this series, the divalent Cu$^{\rm II}$ and Fe$^{\rm
II}$ metal atoms reside decorating and nodal sites of the
diamond-like decorated honeycomb lattice, respectively (see
figure~\ref{fig1}a). Unfortunately, the divalent Fe$^{\rm II}$
atoms are due to a strong ligand field of the cyano group in the
diamagnetic low-spin state with $S = 0$. Similarly, the bimetallic
polymeric coordination compounds
$\rm [W\{(CN)_4Fe(H_2O)_2\}_2]\cdot nH_2O$~\cite{Pil01} and
$\rm [W\{(CN)_4Co(H_2O)_2\}_2]\cdot nH_2O$~\cite{Her03} with the
diamond-like decorated square network structure (see
figure~\ref{fig1}b) have also been reported quite recently, but
the tetravalent W$^{\rm IV}$ metal atoms residing nodal sites of
the diamond-like decorated square lattice are diamagnetic due to a
strong ligand field of the cyano group in those systems.

\section*{Acknowledgements}

This work was partially supported by the internal grant of P.J.~\v{S}af\'arik University
under the contract No. VVGS 1/10--11 and by European Union European regional development fond (ERDF EU) grant under the contract No. ITMS26220120005 (activity 3.2.).

%
%
%

\ukrainianpart

\title{Фазові переходи у геометрично фрустрованій змішаній спін-1/2 і спін-1 моделі Ізінга-Гайзенберга на ромбічноподібних декорованих плоских ґратках}
\author{Л. Ґалісова\refaddr{label1},
        Й. Стречка\refaddr{label2}}
\addresses{
\addr{label1} Механіко-інженерний факультет, Технічний університет, Кошіце, Словацька республіка
\addr{label2} Природничий факультет, Університет ім.~П.Й.~Шафарика, Кошіце, Словацька республіка}

\makeukrtitle

\begin{abstract}
\tolerance=3000%
Досліджуються фазові переходи у змішаній спін-1/2 і спін-1 моделі Ізінга-Гайзенберга на декількох декорованих  плоских ґратках, що складаються зі сполучених ромбів в рамках узагальненого декораційно-ітераційного перетворення. Основна увага приділяється систематичному вивченню фазових діаграм при скінчених температурах в залежності від топології гратки. Критична поведінка гібридної квантово-класичної моделі Ізінга-Гайзенберга порівнюється з  поведінкою її напівкласичного ізінгівського аналога. Показано, що обидві моделі на ромбічноподібних декорованих плоских ґратках демонструють цікаву критичну поведінку, включаючи зворотні фазові переходи. Чим вище координаційне число ґратки, тим помітнішою може бути зворотність.
\keywords модель Ізігнга-Гайзенберга, декораційно-ітераційне перетворення, геометрична фрустрація, зворотні фазові переходи
\end{abstract}


\begin{thebibliography}{99}
\bibitem{Lhu02}
Lhuillier~C., Misguich~G., in: High Magnetic Fields: Applications
in Condensed Matter Physics and Spectroscopy, Edited by
Berthier~C., L\'evy~L.P., Martinez~G., Springer-Verlag, Berlin,
2002;\\
\href{http://dx.doi.org/10.1007/3-540-45649-X_6}
{doi:10.1007/3-540-45649-X$\_$6}
\bibitem{Rich04}
Richter~J., Schulenburg~J., Honecker~A., Quantum Magnetism,
Lecture Notes in Physics, Vol.~645, Edited by Schollw\"ock~U.,
Richter~J., Farnell~D.J.J., Bishop~R.F., Springer-Verlag, Berlin,
2004;\\ \bibdoi{10.1007/BFb0119592}.
\bibitem{Mis04}
Misguich~G., Lhuillier~C., Frustrated Spin Systems, Edited by
Diep~H.T., World Scientific, Singapore, 2004.
\bibitem{Vil77}
Villian~J., J. Phys. C: Solid State Phys., 1977, {\bf 10}, 1717;
\bibdoi{10.1088/0022-3719/10/10/014}.
\bibitem{Vil80}
Villain~J. et al., J. Physique, 1980, {\bf 41}, 1263;
\bibdoi{10.1051/jphys:0198000410110126300}.
\bibitem{Zhi03}
Zhitomirsky~M.E., Phys. Rev. B, 2003, {\bf 67}, 104421;
\bibdoi{10.1103/PhysRevB.67.104421}.
\bibitem{Hon05}
Honecker~A., Richter~J., Condens. Matter Phys., 2005, {\bf 8}, 813.
\bibitem{Der06}
Derzhko~O., Richter~J., Eur. Phys. J. B, 2006, {\bf 52}, 23;
\bibdoi{10.1140/epjb/e2006-00273-y}.
\bibitem{Sch07}
Schnack~J., Schmidt~R., Richter~J., Phys. Rev. B, 2007, {\bf 76},
054413; \bibdoi{10.1103/PhysRevB.76.054413}.
\bibitem{Hon04}
Honecker~A., Schulenburg~J., Richter~J., J. Phys.: Condens.
Matter, 2004, {\bf 16}, S749;\\
\bibdoi{10.1088/0953-8984/16/11/025}.
\bibitem{Str02}
Stre\v{c}ka~J., Ja\v{s}\v{c}ur~M., Phys. Status Solidi B, 2002,
{\bf 233}, R12;\\
%
\href{http://dx.doi.org/10.1002/1521-3951(200210)233:3<R12::AID-PSSB999912>3.0.CO;2-2}
{doi:10.1002/1521-3951(200210)233:3$<$R12::AID-PSSB999912$>$3.0.CO;2-2}
\bibitem{Can04}
\v{C}anov\'a~L., Stre\v{c}ka~J., Ja\v{s}\v{c}ur~M., Czech. J. Phys., 2004, {\bf 54}, D579.
\bibitem{Str06}
Stre\v{c}ka~J., Ja\v{s}\v{c}ur~M., Acta Phys. Slovaca, 2006, {\bf 56}, 65.
\bibitem{Can06}
\v{C}anov\'a~L., Stre\v{c}ka~J., Ja\v{s}\v{c}ur~M., J. Phys.:
Condens. Matter, 2006, {\bf 18}, 4967;\\
\bibdoi{10.1088/0953-8984/18/20/020}.
\bibitem{Can08}
\v{C}anov\'a~L., Stre\v{c}ka~J., Dely~J., Ja\v{s}\v{c}ur~M.,
Acta Phys. Pol., 2008, {\bf 113}, 449.
\bibitem{Jas08}
Ja\v{s}\v{c}ur~M., Stre\v{c}ka~J., \v{C}anov\'a~L., Acta Phys. Pol., 2008, {\bf 113}, 453.
\bibitem{Str09}
Stre\v{c}ka~J., \v{C}anov\'a~L., Lu\v{c}ivjansk\'y~T.,
Ja\v{s}\v{c}ur~M., J. Phys.: Conf. Ser., 2009, {\bf 145}, 012058;
\bibdoi{10.1088/1742-6596/145/1/012058}.
\bibitem{Can09}
\v{C}anov\'a~L., Stre\v{c}ka~J., Lu\v{c}ivjansk\'y~T.,
Condens. Matter Phys., 2009, {\bf 12}, 353.
\bibitem{Can10}
\v{C}anov\'a~L., Stre\v{c}ka~J., Phys. Status Solidi B, 2010, {\bf
247}, 433; \bibdoi{10.1002/pssb.200945444}.
\bibitem{Fis59}
Fisher~M.E., Phys. Rev., 1959, {\bf 113}, 969;
\bibdoi{10.1103/PhysRev.113.969}.
\bibitem{Syo72}
Syozi~I., Phase Transition and Critical Phenomena, Vol. 1, Edited
by Domb~C., Green~M.S., Academic Press, New--York, 1972.
\bibitem{Roj09}
Rojas~O., Valverde~J.S., de Sousa~S.M., Physica A, 2009, {\bf
388}, 1419; \bibdoi{10.1016/j.physa.2008.12.063}.
\bibitem{Rud08a}
Rudowicz~C., Physica B, 2008, {\bf 403}, 1882;
\bibdoi{10.1016/j.physb.2007.10.219}.
\bibitem{Rud08b}
Rudowicz~C., Physica B, 2008, {\bf 403}, 2312;
\bibdoi{10.1016/j.physb.2007.12.011}.
\bibitem{Hou50}
Houtappel~R.M.F., Physica (Amsterdam), 1950, {\bf 16}, 425;
\bibdoi{10.1016/0031-8914(50)90130-3}.
\bibitem{Ons44}
Onsager~L., Phys. Rev., 1944, {\bf 65}, 117;
\bibdoi{10.1103/PhysRev.65.117}.
\bibitem{Bax89}
Baxter~R.J., Choy~T.C., Proc. R. Soc. London, Ser. A, 1989, {\bf
423}, 279; \bibdoi{10.1098/rspa.1989.0055}.
\bibitem{Bar88}
Barry~J.H., Khatun~M., Tanaka~T., Phys. Rev. B, 1988, {\bf 37},
5193; \bibdoi{10.1103/PhysRevB.37.5193}.
\bibitem{Kha90}
Khatun~M., Barry~J.H., Tanaka~T., Phys. Rev. B, 1990, {\bf 42},
4398; \bibdoi{10.1103/PhysRevB.42.4398}.
\bibitem{Bar91}
Barry~J.H., Tanaka~T., Khatun~M., M\'unera~C.H., Phys. Rev. B,
1981, {\bf 44}, 2595;\\ \bibdoi{10.1103/PhysRevB.44.2595}.
\bibitem{Cal63}
Callen~H.B., Phys. Lett., 1963, {\bf 4}, 161;
\bibdoi{10.1016/0031-9163(63)90344-5}.
\bibitem{Suz65}
Suzuki~M., Phys. Lett., 1965, {\bf 19}, 267;
\bibdoi{10.1016/0031-9163(65)90978-9}.
\bibitem{Bal02}
Balcerzak~T., J. Magn. Magn. Mater., 2002, {\bf 246}, 213;
\bibdoi{10.1016/S0304-8853(02)00056-2}.
\bibitem{Diep04}
Diep~H.T., Giacomini~H., Frustrated Spin Systems, Edited by
Diep~H.T., World Scientific, Singapure, 2004.
\bibitem{Hon04b}
Hong~Ch.S., You~Y.S., Inorg. Chim. Acta, 2004, {\bf 357}, 3271;
\bibdoi{10.1016/j.ica.2004.04.004}.
\bibitem{Zha00}
Zhang~H.-X., et al., J. Organomet. Chem., 2000, {\bf 598}, 63;
\bibdoi{10.1016/S0022-328X(99)00679-8}.
\bibitem{Tra01}
Tr\'avni\v{c}ek~Z., Sm\'ekal~Z., Escuer~A., Marek~J., New J.
Chem., 2001, {\bf 25}, 655; \bibdoi{10.1039/b006741p}.
\bibitem{Pil01}
Pilkington~M. et al., J. Solid St. Chem., 2001, {\bf 159}, 262;
\bibdoi{10.1006/jssc.2001.9155}.
\bibitem{Her03}
Herrera~J.M. et al., Inorg. Chem., 2003, {\bf 42}, 7052;
\bibdoi{10.1021/ic034188+}.
\end{thebibliography}
\end{document}